\documentclass[twocolumn,showpacs,preprintnumbers,amsmath,amssymb,floatfix]{revtex4}
\usepackage{graphicx}% Include figure files
\usepackage{dcolumn}% Align table columns on decimal point
\usepackage{bm}% bold math

\begin{document}

\title{Seasonal variation of atmospheric leptons as a probe of charm}

\author{
Paolo Desiati$^{1}$,
Thomas K. Gaisser$^{2}$
}

\affiliation{$^{1}$IceCube Research Center, University of Wisconsin, Madison, WI 53703, U.S.A.}
\affiliation{$^{2}$Bartol Research Institute, University of Delaware, Newark, DE 19716, U.S.A.}

%+\addtocontents{toc}{{\it T. Gaisser}}
%+\label{GaisserStart}

\begin{abstract}

The intensity of TeV atmospheric muons and neutrinos
depends on the temperature in the stratosphere.  We show that the energy-dependence
in the 100 TeV range of the correlation with temperature is sensitive to the fraction 
of muons and neutrinos from decay of charmed hadrons.  We discuss the prospects
for using the temperature effect as observed in gigaton neutrino detectors
to measure the charm contribution.
\end{abstract}

%\pacs{14.60.Ef, 14.60.Lm, 14.60.St, 14.20.Lq, 92.60.hv, 92.60.hd}
\maketitle

Atmospheric muons and neutrinos are important for neutrino telescopes, both as
background in the search for high-energy neutrinos of astrophysical origin
and because they serve to calibrate the detectors.  In addition, the atmospheric 
lepton spectra are of interest because the large size of 
current and proposed neutrino telescopes will make it possible to extend
the measurements of atmospheric muons and neutrinos beyond the reach of previous
measurements.  It may finally be possible to achieve a long-sought goal of cosmic-ray
science by detecting the contribution to the atmospheric leptons from decay of
charmed hadrons, the ``prompt" neutrinos and muons.  At energies much lower than
$\sim 100$ TeV, the prompt component is hidden by
the much more abundant contributions from decay of charged pions and kaons.
A temperature dependent signal of the charm contribution would complement
the standard approach of looking for an isotropic component of the muon flux
with a harder spectrum.

It has long been known that the intensity of high-energy muons depends on
the temperature in the stratosphere~\cite{Barrett}.  The variation arises
from the corresponding expansion and contraction of the atmosphere at high 
altitude where most production of the muons occurs.  A similar effect is also
expected for neutrinos for the same reason~\cite{AckermannBernardini}.  The
temperature dependence of TeV muons measured in MINOS~\cite{MINOS} has recently been used
as a probe of sudden stratospheric warmings~\cite{GDBarr}.  IceCube reported
a correlation of muon intensity with stratospheric temperature with 
high precision made possible by the high statistics of the kilometer-scale 
detector~\cite{Tilavetal}.
Ref.~\cite{Grashorn} describes how the temperature dependence
of muon intensity in the 1 to 10 TeV range
is sensitive to the kaon to pion ratio in the atmosphere.  In this
paper we point out that kilometer-scale neutrino detectors may have sufficient
statistics in the 100 TeV range to be able to constrain the contribution of
charm decay to prompt leptons.

Currently operating neutrino telescopes are Baikal~\cite{Baikal},
Antares~\cite{Antares} and IceCube~\cite{IceCube}.  AMANDA~\cite{Kelley},
the predecessor of IceCube at the South Pole, was turned off at the
beginning of 2009.  With a planned volume of one cubic kilometer, IceCube
will be by far the largest operating neutrino detector.  It is
already running with 90\% of its sensors installed
and will be completed in 2011.  There are plans for
a still larger neutrino telescope in the Mediterranean Sea (Km3NeT)~\cite{Km3NeT},
and there are also plans to expand the Baikal detector~\cite{GVD}.

Preliminary measurements of muon~\cite{Berghaus} and neutrino~\cite{Chirkin}
spectra with data from the partially constructed (25\% complete) IceCube
in 2007 were reported at the
International Cosmic Ray Conference, July 2009.  The atmospheric
neutrino spectrum measured with AMANDA with data taken from 2000 to
2006~\cite{Kelley} already extends to just above {30}~{TeV} with high statistics.
An unfolding analysis of the AMANDA~\cite{Becker} data extends the reach of the
same data set to above 100~TeV.  No contribution of a prompt component is
yet visible in the unfolded neutrino spectrum.  The full
IceCube will be large enough to extend measurements of atmospheric 
neutrinos well beyond 100~TeV with good statistics.  From the point of view
of precise analysis of the energy dependence of the correlation coefficient,
the extremely high statistics available with muons in a kilometer scale
detector is also important.

Fluxes of secondary cosmic-ray leptons in the atmosphere can be described to a good
approximation by a set of formulas in which each element corresponds to one of
the processes involved in their production~\cite{Gaisser,Lipari}.  
For $\nu_\mu + \bar{\nu}_\mu$
\begin{eqnarray}
&\phi_\nu(E_\nu)& =  \phi_N(E_\nu) \times \nonumber \\
  & & \left\{{A_{\pi\nu}\over 1 + 
B_{\pi\nu}\cos\theta\, E_\nu / \epsilon_\pi}
\,+\,{A_{K\nu}\over 1+B_{K\nu}\cos\theta\, E_\nu / \epsilon_K}\right.\nonumber \\
& & \left. +\,\,\,{A_{{\rm charm}\,\nu}\over 1+B_{{\rm charm}\,\nu}\cos\theta\, E_\nu / \epsilon_{\rm charm}}\right\},
\label{angular}
\end{eqnarray}
where $\phi_N(E_\nu) = dN/d\ln(E_\nu)$ is the primary spectrum
of nucleons ($N$) evaluated at the energy of the neutrino.
The three terms in brackets correspond to production from leptonic
and semi-leptonic decays of pions, kaons and charmed hadrons respectively.  
The equation for
muons is similar at energies sufficiently high ($>100$~GeV) so that energy loss
and decay of muons in the atmosphere can be neglected.  The numerator of each
term is of the form
\begin{equation}
A_{i\nu}\;=\;{Z_{i\nu}\times BR_{i\nu} \times Z_{i\nu} \over 1-Z_{NN}}
\label{numerator}
\end{equation}
with $i\,=\,\pi^\pm,\,K,\,{\rm charm}$ and $BR_{i\nu}$ is the branching ratio
for $i\rightarrow\nu$.   The first $Z$-factor in the numerator 
is the spectrum weighted moment
of the cross section for a nucleon (N) to produce a secondary hadron $i$
from a target nucleus in the atmosphere, and the second $Z$-factor is the
corresponding moment of the decay distribution for $i\rightarrow \nu + X$.
Thus, for example for $N\,+\,{\rm air}\,\rightarrow\,K^+\,+\,X$,
\begin{equation}
 Z_{NK^+}\;=\;{1\over \sigma_{N-{\rm air}}}\int_0^1\,x^\gamma{{\rm d}\sigma_{NK^+}(x)\over{\rm d}x},
\label{Zfactor}
\end{equation}
where $x\,=\,E_{K^+}/E_N$ and $\gamma\approx 1.7$ is the integral spectral index of the
incident spectrum of cosmic-ray nucleons.  The denominator of Eq.~\ref{numerator}
is the ratio of nucleon interaction to attenuation length.  For calculations
below we use numerical values of the parameters from Ref.~\cite{Gaisser}.  Parameters
for charm production are discussed below.

The denominator of each term in Eq.~\ref{angular} reflects an important physical property
of meson decay in the atmosphere--the competition between decay and
reinteraction of hadrons.  The critical energy depends on the zenith angle $\theta$
of the cascade in the atmosphere, and is of the form
\begin{equation}
 E_{\rm critical}\;=\;{\epsilon_i\over \cos{\theta^*}}\,=\,{m_ic^2h_0\over\cos{\theta^*}\,c\tau_i},
\label{critical}
\end{equation}
where $\theta^*$ is the local zenith angle at lepton production taking account
of the curvature of the Earth~\cite{Lipari}.
The $\epsilon_i$ are characteristic energies for each channel and
$h_0\approx 6.4$~km is the scale height in an exponential approximation to
the density of the atmosphere at high altitude.  When $E_i\,<\,\epsilon_i/\cos{\theta}$
the mesons decay so that the low-energy neutrino spectrum has the same power law index
as the primary cosmic-ray spectrum.  At high energy, the contribution of each term
gradually steepens so that asymtotically, the neutrino spectrum is one power steeper
than the primary spectrum.  The characteristic energies are
\begin{center}
$\epsilon_\pi=115$~GeV,  $\epsilon_{K^\pm}=850$~GeV, and $\epsilon_{\rm charm}\sim 5\times 10^7$~GeV.
\end{center}

These differences lead to a characteristic pattern of the contributions of the
various channels to the flux of neutrinos: first the pion contribution steepens,
then the kaon contribution and finally (at much higher energy) the contribution 
from decay of charmed hadrons becomes the main source of
neutrinos and muons in the atmosphere.  In addition, the suppression of the charged pion
contribution to the neutrino flux starts early because the quantity $B_{\pi\nu}=2.8$
in the denominator of the first term of Eq.~\ref{angular} is anomalously large
compared to the corresponding factors for strange and charmed hadrons.  This is a 
consequence of the kinematics of $\pi^\pm\rightarrow\mu^\pm\,+\,\nu$ decay in flight 
in which the muon carries most of the energy because its mass is 
comparable to that of the parent pion.

%%%%%%%%%%%%%%%%%%%%%%%%%%%%%%%%%%%%%%%%%%%%%%%%%%%%%%%%%%%%%%%%%%%%%%%%%
%%
%%   use this format to include an .eps figure into your paper
%%
\begin{figure}[htb]
\begin{center}
\includegraphics[width=2.5in]{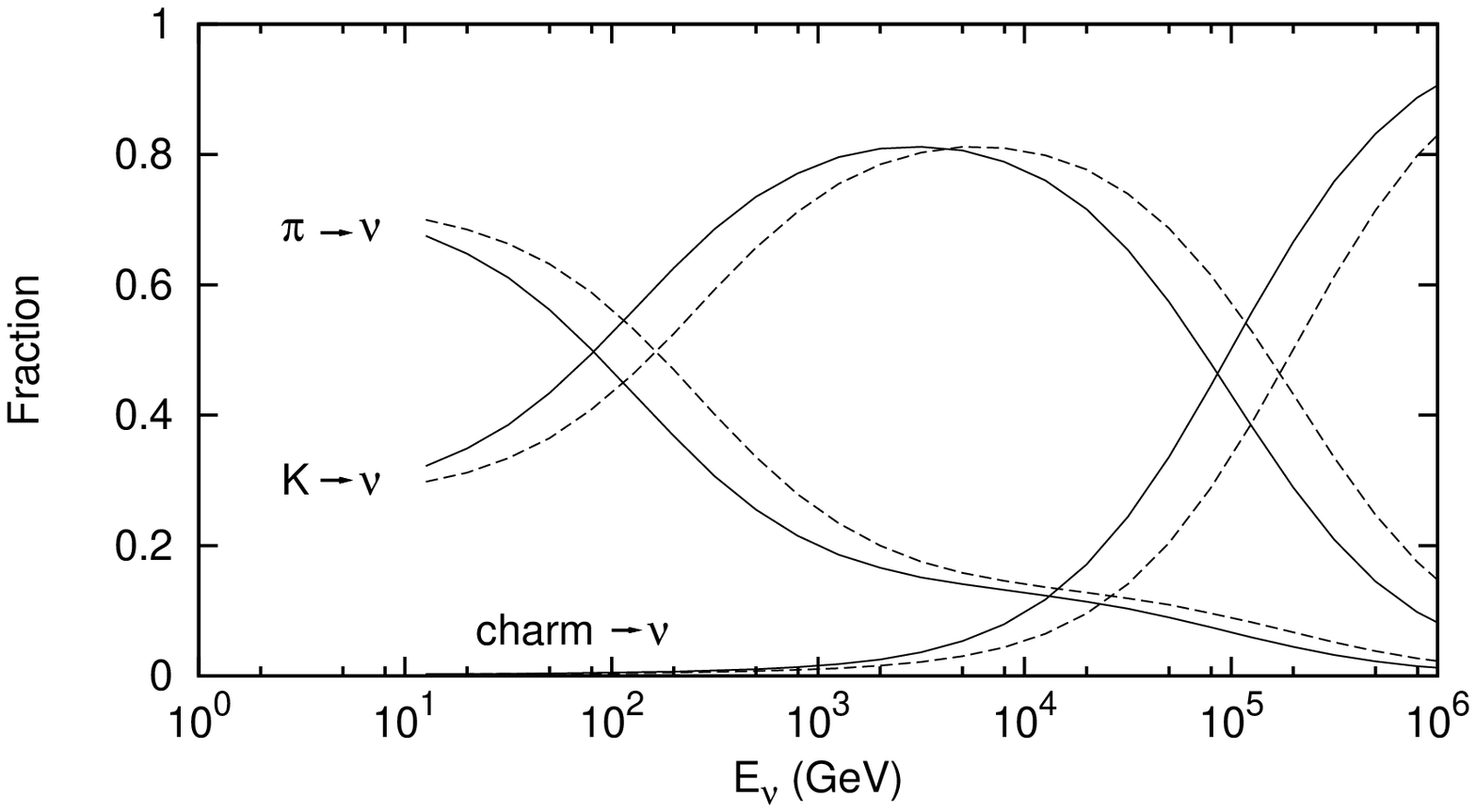} \\
\includegraphics[width=2.5in]{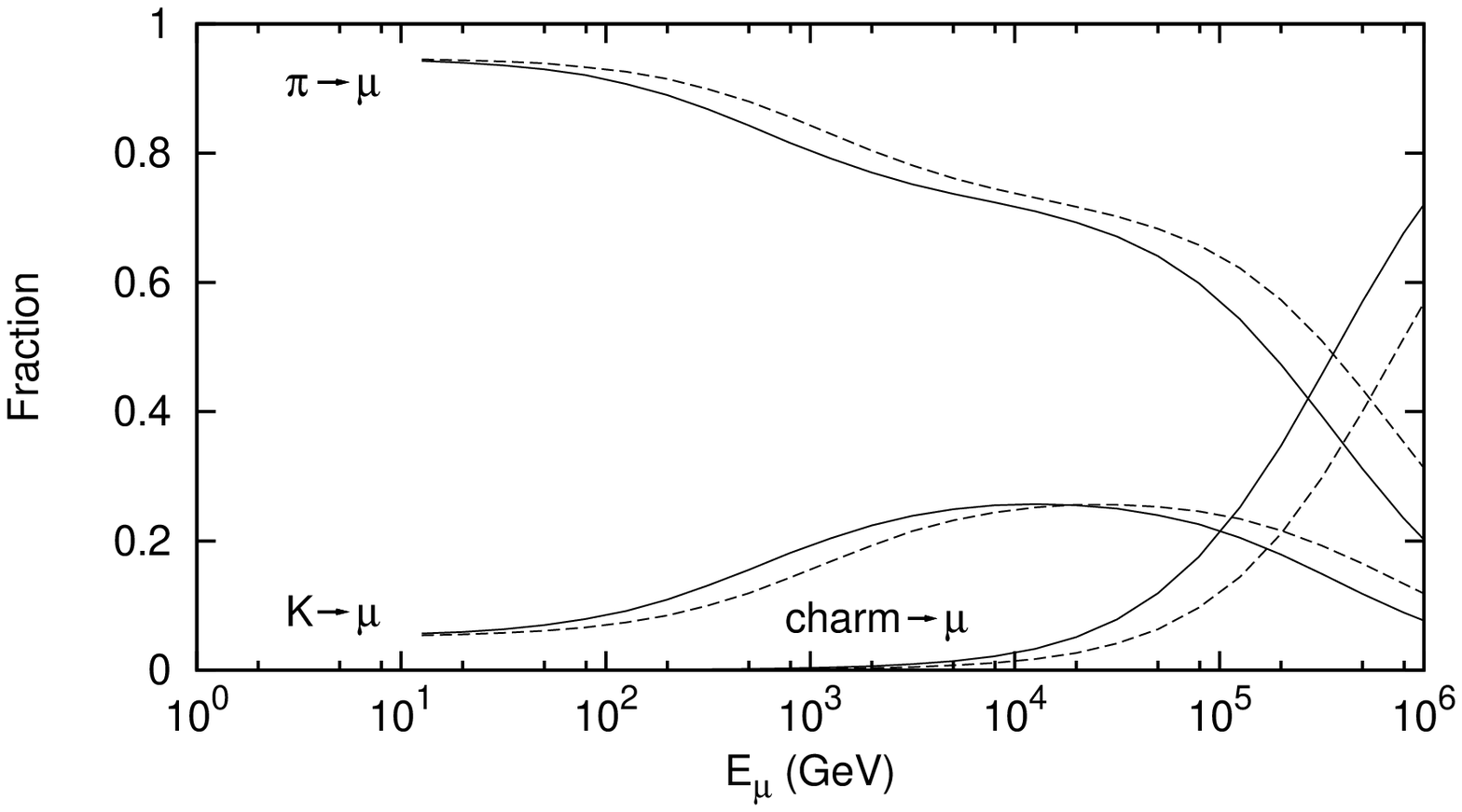}
\caption{Top: fraction of muon-neutrinos from pion decay, kaon decay
and charm decay (RQPM~\cite{Bugaev}) as a function of neutrino energy; Bottom: same for muons.  Solid 
lines are for vertical and dashed lines for $60^\circ$ (see text for discussion).}
\label{fig:ratio}
\end{center}
\end{figure}
%%%%%%%%%%%%%%%%%%%%%%%%%%%%%%%%%%%%%%%%%%%%%%%%%%%%%%%%%%%%%%%%%%%%%%%%%%%

Fig.~\ref{fig:ratio} shows the fractional contribution of the main hadronic channels
to the production of neutrinos and muons in the atmosphere.
Kaons actually become the dominant source of neutrinos above $\sim100$~GeV.  For
muons, the dominant channel is never kaons, but the contribution of kaons does
increase significantly in the TeV energy range.  An interesting manifestation 
of the increasing importance of kaons at high energy is the measurement of
the charge ratio of atmospheric muons with energies at production 
in the TeV range~\cite{MINOSchargeratio,OPERA}.  The 
$\mu^+/\mu^-$~ratio is observed to increase
from its value of 1.27 around $100$~GeV to $\approx 1.37$ above a TeV.  The increase is
attributed to the importance of the associated production of kaons,
\begin{equation}
 p\,+\,air\,\rightarrow\,\Lambda\, +\, K^+\, +\, X,
\label{associated}
\end{equation}
which makes the $+/-$ charge ratio larger for kaons than for pions.
Production of strange particles and anti-particles--in particular the process 
in Eq.~\ref{associated}--is highly asymmetric in the forward fragmentation region.  Production of
$\Lambda$ is favored because it has constituents in common with the
valence quarks of the proton.

At some energy, perhaps
around $100$~TeV or somewhat above, the decay of charmed hadrons will become
the main source of all atmospheric leptons.  Where the transition occurs
is uncertain because of the wide variation in the literature on the level of charm
production at large Feynman x.  Like associated production of strangeness via
Eq.~\ref{associated}, the production of $\Lambda_c^+$ is also highly asymmetric
in the forward fragmentation region.  A measurement by the Fermilab Experiment 781 (SELEX)~\cite{SELEX}
shows that $(\sigma_c\,-\,\sigma_{\bar c})\,/\,(\sigma_c\,+\,\sigma_{\bar c})\,\approx\,1$
for $0.2 < x < 0.7$ for $600$~GeV protons on a fixed target.
Here $\sigma_c$ and $\sigma_{\bar c}$ represent respectively the production
of $\Lambda_c^+$, which does, and $\bar{\Lambda}_{\bar c}^-$, which does not
have valence quarks in common with the beam proton.  A large contribution from
the charmed analog process to Eq.~\ref{associated} could be classified as
"intrinsic" charm~\cite{Brodsky}.  There is some support for a component of
intrinsic charm from recent measurements of charm production on different
nuclear targets~\cite{SELEX2}, as discussed in Ref.~\cite{Kop}.

The RQPM model of charm production~\cite{Bugaev}
includes such a contribution and predicts a relatively high level of prompt leptons, 
as illustrated in Fig.~\ref{fig:ratio}.
Calculations using perturbative QCD tend to give lower
levels of charm production.  As an example, a recent calculation within a perturbative QCD framework~\cite{Enberg} predicts a contribution from charm decay roughly an order of
magnitude lower than the RQPM model.
In what follows we
evaluate the sensitivity of seasonal effects to prompt atmospheric leptons
for three different assumptions about the level of charm production.  
We compare RQPM with charm production at the level of Ref.~\cite{Enberg} (henceforth ERS)
and with a somewhat arbitrary intermediate model
described in terms of the parameters of Eq.~\ref{numerator}.  Representing
the weighted sums of the various charm channels and their semi-leptonic decay
modes with
$Z_{\rm N-charm} = 5\times 10^{-4}$, $Z_{{\rm charm}\nu} = Z_{{\rm charm}\mu} = 0.13$
and an effective semi-leptonic branching fraction for charm decays of $0.14$, gives
a level of charm production about a factor of two lower than RQPM.

\begin{figure}[htb]
\begin{center}
\includegraphics[width=2.5in]{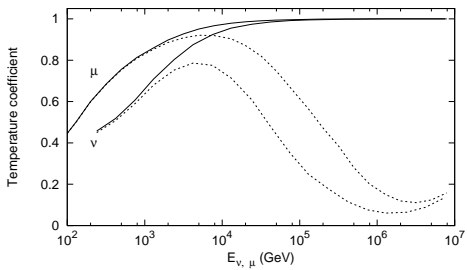}
\caption{Differential correlation coefficient
as a function of lepton energy for near vertical 
muons and muon neutrinos.  Solid lines: no charm;
dashed lines: with RQPM charm~\cite{Bugaev}.} 
\label{alpha-diff}
\end{center}
\end{figure}

In an isothermal approximation, the density  of the atmosphere is described
by an exponential with a scale height of $h_0\approx 6.4$~km, where the numerical
value is applicable to the stratosphere where most high-energy muons and neutrinos
originate.  From the ideal gas equation relating density and pressure, one finds
$h_0\;=\;R\,T$.
With this relation it is then possible from Eqs.~\ref{critical} and~\ref{angular}
to calculate the variation of the flux with temperature.  The deviation of
the atmosphere from isothermal is accounted for to first order in $\delta T / T$
by weighting the actual temperature profile by the profile of production of the
atmospheric leptons.  It is conventional
to define a correlation coefficient, $\alpha_\mu$ which relates the fractional change
in the atmospheric muon intensity to the fractional change in the temperature.
One can describe the atmospheric neutrinos
in parallel, with a coefficient $\alpha_\nu$~\cite{AckermannBernardini}.
In differential form the correlation coefficient is
%\begin{equation}
%{1\over \phi(E,\theta)}{{\rm d}\phi(E,\theta)\over dT}\delta T\;=
%\;\alpha(E,\theta)\times{\delta T\over T},
%\end{equation}
%so that
\begin{equation}
\alpha(E,\theta)\;=\;T\,{1\over \phi(E,\theta)}{{\rm d}\phi(E,\theta)\over dT}.
\label{diff-alpha}
\end{equation}
The derivative can be calulated directly from the expressions \ref{angular}
and \ref{critical}, with the results shown in Fig.~\ref{alpha-diff}.

The features of the curves in Fig.~\ref{alpha-diff} correspond to the
properties of the various terms in Eq.~\ref{angular}.  At extremely high
energy, the second term in the denominator dominates, which means the
the intensity of muons is proportional to temperature and the correlation
coefficient is unity.  Asymptopia is reached at lower energy for pions than for kaons,
so the inclusion of kaons delays the increase of the correlation coefficient
so somewhat higher energy.  This is more apparent for neutrinos than for muons
because kaons give a larger fractional contribution to neutrinos than to muons.
The increase of the correlation
coefficient is delayed to higher energy
at large zenith angle because the critical energy (Eq.~\ref{critical}) is larger.
Since the critical energy for the charm contribution is at extremely high
energy ($\sim 5\times10^7$~GeV), the prompt leptons from charm decay
have no temperature dependence until $\sim 10^7$~GeV.  Thus, a significant
contribution from charm will suppress the correlation coefficient above an
energy that depends on the magnitude of the charm contribution to 
lepton production.  For the example shown in Fig.~\ref{alpha-diff} in which the
RQPM model of Bugaev et al.~\cite{Bugaev} is used, the suppression begins
already between 10 and 100 TeV depending on zenith angle.

Measured rates depend on the convolution of the lepton spectrum with the
detector response and effective area, which depends on lepton energy and direction.
To compare with measurements, it is 
therefore necessary to make the following calculation:
\begin{eqnarray}
\alpha_i(\theta)&=&{T\over \int {\rm d}E\,\phi_i(E)\times A_{i,{\rm eff}}(E,\theta)}
 \nonumber \\
&\times & {{\rm d}\over {\rm d}T}\,\int {\rm d}E\,\phi_i(E)\times A_{i,{\rm eff}}(E,\theta).
\label{integral}
\end{eqnarray}
Here the subscript $i = \nu, \mu$ and $A_{i,{\rm eff}}(E,\theta)$ is the energy-dependent
effective area of the detector as seen from a direction $\theta$.  
In the remainder of the paper we explore some examples relevant for IceCube.

To make an estimate for muons
we assume a step function at energies corresponding to the minimum needed
to penetrate through IceCube treated as a sphere with a projected
area of one square kilometer.  
Table~\ref{Table1} gives some rates for muons with zenith angle $<\,30^\circ$.
The correlation coefficient is $10$\% lower with RQPM charm
for $E_\mu\,>10\,$~TeV than with no charm and the rate is more 
than one Hz per square kilometer.  For $E_\mu\,>\,100$~TeV
the correlation coefficient is only $0.5$ with charm while the event
rate is much lower, but still of order $10^5$ events per year
in a kilometer scale detector.

\begin{table}
\begin{tabular}{|l||l|r||l|r||l|r||l|r|} \hline\hline
{$E_{\mu,{\rm min}}$} &
\multicolumn{2}{c||} {no charm} &
\multicolumn{2}{c||} {RQPM charm} & 
\multicolumn{2}{c||} {ERS charm} &
\multicolumn{2}{c|} {int. charm} \\ \hline
 & $\alpha$ & Rate & $\alpha$ & Rate &
$\alpha$ & Rate & $\alpha$ & Rate \\ \hline
0.5 & 0.83 & 2050 & 0.82 & 2070 & 0.82 & 2050 & 0.82 & 2060\\ \hline
10 & 0.98 & 1.26 & 0.89 & 1.40 & 0.97 & 1.26 & 0.94 & 1.34 \\ \hline
100 & 1.0 & 0.0025 & 0.53 & 0.0049 & 0.91 & 0.0028 & 0.71 & 0.0036 \\ \hline\hline
\end{tabular}
\caption{\label{Table1}{Correlation coefficients for muons with
($\theta\,\le\,30^\circ$) for three levels of charm 
(energy in TeV; rate in Hz/km$^2$).}}
\end{table}

For neutrinos we use
the effective areas computed for IceCube-40~\cite{DummLodz}
multiplied by a factor of two to estimate the event numbers
for a full cubic kilometer detector.  Table~\ref{Table2} gives correlation
coefficient and expected rates for three different ranges of
zenith angle for neutrino-induced muons coming into the detector
from below.  The three ranges of zenith angle are 1) $90^\circ \,< \theta\,< 120^\circ$,
2) $120^\circ\,<\,\theta\,<\,150^\circ$, and 3) $150^\circ\,<\,\theta\,<\,180^\circ$,
which have solid angles respectively of $\pi$, $0.73\pi$ and $0.27\pi$.
For a detector at the South Pole, zone 1 corresponds approximately to the 
Southern temperate atmosphere, zone 2 to the tropics and zone 3 to the Northern
temperate atmosphere.  
Absorption of neutrinos in the Earth is significant for the high-energy
entries in the vertical bins.

\begin{table}
\begin{tabular}{|l||l|r||l|r|} \hline\hline
{$E_{\nu,{\rm min}}$}(TeV) &
\multicolumn{2}{c||} {no charm} &
\multicolumn{2}{c|} {RQPM charm} \\ \hline
Zone 1 & $\alpha$ & Events/yr  & $\alpha$ & Events/yr  \\ \hline
all & 0.54 & 16000 & 0.52 & 17000 \\ \hline
3 & 0.70 & 5900 & 0.62 & 6300 \\ \hline
30 & 0.94 & 350 & 0.72 & 450 \\ \hline\hline
\end{tabular}

\begin{tabular}{|l||l|r||l|r|} \hline\hline
{$E_{\nu,{\rm min}}$}(TeV) &
\multicolumn{2}{c||} {no charm} &
\multicolumn{2}{c|} {RQPM charm} \\ \hline
Zone 2 & $\alpha$ & Events/yr & $\alpha$ & Events/yr \\ \hline
all & 0.66 & 6000 & 0.62 & 6400 \\ \hline
3 & 0.88 & 1230 & 0.75 & 1450 \\ \hline
30 & 0.98 & 37 & 0.46 & 80 \\ \hline\hline
\end{tabular}

\begin{tabular}{|l||l|r||l|r|} \hline\hline
{$E_{\nu,{\rm min}}$}(TeV) &
\multicolumn{2}{c||} {no charm} &
\multicolumn{2}{c|} {RQPM charm} \\ \hline
Zone 3 & $\alpha$ & Events/yr  & $\alpha$ & Events/yr \\ \hline
all & 0.68 & 1650 & 0.64 & 1750 \\ \hline
3 & 0.91 & 260 & 0.75 & 320 \\ \hline
30 & 0.99 & 5.2 & 0.41 & 13 \\ \hline\hline
\end{tabular}
\caption{\label{Table2}
{Correlation coefficients with and without charm for neutrinos in 
three zones of the atmosphere (see text).}}
\end{table}

For lepton energies below 10 PeV the contribution from charm decay
does not depend on temperature because charmed hadrons decay before
interacting as a conseqence of their short decay times.  In contrast,
by 100 TeV pions and kaons are fully asmyptotic in the sense that
their decay probability is proportional to temperature.
This situation 
leads to the possibility of inferring the level of
prompt lepton production from a precise measurement of the correlation
of the rate of leptons with stratospheric temperature
in the 10 to 100 TeV energy range.  There are several
practical problems that must be overcome to realize this possibility.
First, a high-energy event sample must be defined with sufficient
energy resolution and sufficient statistics to see the effect of the
charm contribution on the correlation with temperature.
In the case of muons, there will be sufficient statistics in a
kilometer scale detector, but the energy assignment of individual muons
may be complicated by the presence of accompanying muons from the same
cosmic-ray primary.
In the case of neutrinos it will be necessary to integrate over a large fraction of
the surface of the Earth, although about 25\% of up-ward atmospheric neutrinos
are generated in the atmosphere above Antarctica, where the stratospheric temperature variation is strongest. The practical application of this
probe of charm is being developed in association with
analysis of measured temperature effects in IceCube
and is most promising for high-energy muons because of
the large statistics.
%
%In the case of neutrinos it will be necessary to integrate
%over a large fraction of the surface of the Earth to accumulate a sufficiently
%large sample of events, making it more difficult to determine the effective
%temperature for the sample.
%Nevertheless, the correlation with stratospheric temperature
%should provide a useful extra handle on distinguishing the prompt lepton component
%in kilometer scale detectors.

ACKNOWLEDGMENTS. PD acknowledges the support from the U.S. National Science Foundation-Office of Polar Programs. TKG is grateful for a Humboldt Foundation Research Prize
and for hospitality at DESY-Zeuthen, the University of Wuppertal and the Karlsruhe Institute of Technology during the period
of this research, which is also supported in part by NSF Award 0856253.


\begin{thebibliography}{99}

%%
%%  bibliographic items can be constructed using the LaTeX format in SPIRES:
%%    see    http://www.slac.stanford.edu/spires/hep/latex.html
%%  SPIRES will also supply the CITATION line information; please include it.
%%
\bibitem{Barrett} P.H. Barrett {\it et al.}, Revs. Mod. Phys. {\bf 24}, 133 (1952).
\bibitem{AckermannBernardini} M. Ackermann \& E. Bernardini, Proc. 29th Int.
Cosmic Ray Conf. (Pune, India) vol. 9, 107-110 (2007).
\bibitem{MINOS} P. Adamson {\it et al.}, Phys. Rev. D {\bf 81}, 012001 (2010).
\bibitem{GDBarr} S. Osprey {\it et al.}, Geophys. Res. Lett. {\bf 36} L05809 (2009).
\bibitem{Tilavetal} S. Tilav {\it et al.}, Proc. 31st Int. Cosmic Ray Conf. (\L\'od\'z, Poland) paper 1565 (2009) (arXiv:1001.0776).
\bibitem{Grashorn} E. Grashorn, Astropart. Phys. {\bf 33}, 140 (2010).
\bibitem{Baikal} V. Aynutdinov {\it et al.}, Astropart. Phys. 25, 140 (2006).
\bibitem{Antares} J.A. Aguilar {\it et al.}, Nucl. Inst. Methods A 570, 170 (2007).
\bibitem{IceCube} A. Achterberg {\it et al.}, Astropart. Phys. 26, 155 (2006);
A. Karle {\it et al.}, Proc. 31st Int. Cosmic Ray Conf. (\L\'od\'z, Poland) paper 1339 (2009).
\bibitem{Kelley} R. Abbasi {\it et al.}, Phys. Rev. D79, 102005 (2009) 
\bibitem{Km3NeT} Els de Wolf, Nucl. Inst. Methods A 588, 86 (2008).
\bibitem{GVD} V. Aynutdinov {\it et al.}, Nucl. Inst. Methods A 602, 14 (2009).
\bibitem{Berghaus} P. Berghaus {\it et al.}, Proc. 31st Int.
Cosmic Ray Conf. (\L\'od\'z, Poland) paper 1565 (2009) (arXiv:0909.0679).
\bibitem{Chirkin} D. Chirkin {\it et al.}, Proc. 31st Int.
Cosmic Ray Conf. (\L\'od\'z, Poland) paper 1418 (2009).
\bibitem{Becker} R. Abbasi {\it et al.}, Astropart. Phys. 34, 48 (2010).
\bibitem{Gaisser} T.K. Gaisser, {\it Cosmic Rays and Particle Physics}, (Cambridge University Press, 1990).
\bibitem{Lipari} P. Lipari, Astropart. Phys. 1, 195 (1993).
\bibitem{MINOSchargeratio} P. Adamson {\it et al.}, Phys. Rev. D76 (2007) 052003 and P.A. Schreiner,  J. Reichenbacher, M.C. Goodman, Astropart. Phys. 32, 61 (2009). 
\bibitem{OPERA} N. Agafonova {\it et al.}, arXiv:1003.1907.
\bibitem{SELEX} F.G. Garcia {\it et al.}, Phys. Lett. B528, 49 (2002). 
\bibitem{Brodsky} S.J. Brodsky, P. Hoyer, C. Peterson \& N. Sakai, Phys. Lett. 93B, 451 (1980). 
\bibitem{SELEX2} A. Blanco-Covrrubias {\it et al.}, Eur. Phys. J. C64, 637 (2009).
\bibitem{Kop} B.Z. Kopeliovich, I.K. Potashnikova and I. Schmidt, arXiv:1003.3673.
\bibitem{Bugaev} E.V. Bugaev {\it et al.}, Phys. Rev. D58, 054001 (1998).
\bibitem{Enberg} R. Enberg, M.H. Reno \& I. Sarcevic, Phys. Rev. D78, 043005 (2008).
\bibitem{DummLodz} J. Dumm, {\it et al.}, Proc. 31st Int. Cosmic Ray Conf. (\L\'od\'z, Poland) paper 0653 (2009).

\end{thebibliography}
\end{document}